\documentclass[aps, pra, twocolumn, reprint, final, showpacs, superscriptaddress, floatfix, amssymb, amsmath, amsthm, hidelinks, numbers, sort&compress, floats]{revtex4-1}

\usepackage{epsfig}
\usepackage{graphicx}
\usepackage[normalem]{ulem}
\usepackage{xcolor}
\usepackage{times, graphics, dcolumn}
\usepackage{units}
\usepackage[colorlinks=true,citecolor=blue,allcolors=blue]{hyperref}  
\usepackage{bm, bbm}  
\usepackage[latin1]{inputenc}
\usepackage{ae,aecompl} 

\begin{document}

\title{Matter wave speckle observed in an out-of-equilibrium quantum fluid}

\author{Pedro~E.~S.~Tavares}\email{pedro.ernesto.tavares@usp.br}
\affiliation{Instituto de F\'{i}sica de S\~ao Carlos, Universidade de S\~ao Paulo, Caixa Postal 369, 13560-970, S\~ao Carlos, SP, Brazil}

\author{Amilson~R.~Fritsch}
\affiliation{Instituto de F\'{i}sica de S\~ao Carlos, Universidade de S\~ao Paulo, Caixa Postal 369, 13560-970, S\~ao Carlos, SP, Brazil}

\author{Gustavo~D.~Telles}
\affiliation{Instituto de F\'{i}sica de S\~ao Carlos, Universidade de S\~ao Paulo, Caixa Postal 369, 13560-970, S\~ao Carlos, SP, Brazil}

\author{Mahir~S.~Hussein}
\affiliation{Departamento de F\'{\i}sica, Instituto Tecnol\'{o}gico de Aerona\'{u}tica, DCTA, 12.228-900 S\~ao Jos\'{e} dos Campos, SP, Brazil}

\author{Fran\c{c}ois~Impens}
\affiliation{Instituto de F\'{\i}sica, Universidade Federal do Rio de Janeiro, Caixa Postal 68528, 21941-972 Rio de Janeiro, RJ, Brazil}

\author{Robin~Kaiser}
\affiliation{Universit\'{e} Nice Sophia Antipolis, Institut Non-Lin\'{e}aire de Nice, \\ CNRS UMR 7335, F-06560 Valbonne, France}

\author{Vanderlei~S.~Bagnato}\email{vander@ifsc.usp.br}
\affiliation{Instituto de F\'{i}sica de S\~ao Carlos, Universidade de S\~ao Paulo, Caixa Postal 369, 13560-970, S\~ao Carlos, SP, Brazil}

\date{November 28, 2017}

\begin{abstract}
We report the results of a direct comparison of a freely expanding turbulent Bose-Einstein condensate and the propagation of an optical speckle pattern. We found remarkably similar statistical properties underlying the spatial propagation of both phenomena. The calculated second-order correlation together with the typical correlation length of each system is used to compare and substantiate our observations. We believe that the close analogy existing in between an expanding turbulent quantum gas and a traveling optical speckle, might burgeon into an exciting new research field investigating disordered quantum matter.
\end{abstract}

\maketitle

Coherent matter wave systems such as a Bose-Einstein condensate (BEC) or atom lasers, and atom optics are reality and relevant research fields. The spatial propagation of coherent matter waves has been an interesting topic for many theoretical \cite{Castin1996,Impens2009,Riou2008} and experimental studies \cite{Le2001, Busch2002, Riou2006, Couvert2008}. On the other hand, optical speckle phenomena emerging as the interference of multiple independent wave fronts originated from a rough surface has been around for many years with ever growing interest~\cite{Goodman2010Book}. The generalization of optical speckle to coherent matter waves is not yet well established, though it would be helpful to be able to use the solid statistical framework, developed over decades, in the atom optics. It may certainly create new perspectives and bring a broader and deeper understanding to the matter-wave community.
	Previously, there had been some effort to investigate this topic. For instance, an atomic beam was used to generate speckle in a de Broglie wave~\cite{Patton2006}. In that case, a multi-mode guidance was used together with a time-dependent second-order correlation function to characterize a matter-wave speckle. Also, Dall and co-authors~\cite{Dall2011} produced an atomic speckle by transmitting atoms through an optical waveguide when a multi-mode matter-wave guidance was on, in very close analogy with multi-mode light guiding into optical fibers. 
	In this work, we evaluated the second-order correlation of an experimental turbulent BEC and determined its typical correlation length. The results were then properly compared to the correlation length of a computer generated, elliptical speckle light map, propagating along the $z$ direction, determined from its intensity correlation. We then discuss the remarkable similarities existing in between light and matter wave propagation in the presence of spatial disorder.

	The experimental sequence to produce the turbulent BEC runs as follows. $^{87}\rm{Rb}$ standard BECs containing about $1-2 \times 10^{5}$ atoms in the |$F=2,m_{F}=+2$> hyperfine state, with a small thermal fraction ($\leq 35\%$) in a cigar-shaped Quadrupole-Ioffe configuration (QUIC) magnetic trap, well into the Thomas-Fermi regime ($T\approx170\unit{nK}, T/T_0\approx0.47$). The measured trapping frequencies are: $\omega_z=2\pi \times 21.1(1)\unit{Hz}$ in the symmetry axis, and $\omega_r=2\pi \times 188.2(3)\unit{Hz}$ in the radial direction. In order to drive the BEC towards turbulence we superpose an additional sinusoidal (time-dependent) magnetic field gradient, right after the BEC is produced. An anti-Helmholtz coil pair set with symmetry axis slightly tilted compared to the QUIC trap symmetry axis ($\theta \leq 5^{\circ}$). The field gradient defining the excitation amplitude ranged from 80 to $790\unit{mG/cm}$ along the tilted axis, and a minimum of about $350\unit{mG/cm}$ is needed to drive the BECs into turbulence. The extra gradient oscillates at $189\unit{Hz}$ frequency for 6 full periods ($31.7\unit{ms}$), after which it is completely shut-off. The perturbed BEC is held in the trap for additional $35\unit{ms}$, after which it is switched off and the BEC is let to expand and fall. Resonant, time-of-flight (TOF) absorption images are acquired after variable time lapses in free fall. We checked that the BEC reached the turbulent state by running the following experimental tests. First, we analyze the density fluctuations appearing on the absorption images. Second, we determine the momentum distribution spectrum from the time-of-flight images \cite{Thompson2014,Tsatsos2016} and check the power-law dependence, $n(k)\propto k^{-3}$, which characterizes the existence of an energy cascade, a fundamental feature of turbulence. Though the full characterization of turbulent regimes is still under investigation in our laboratory. It seems to depend on the condensed fraction, but not on the exact excitation conditions, provided that a stationary state was established.
	
\begin{figure}
\centering
\includegraphics[width=\linewidth]{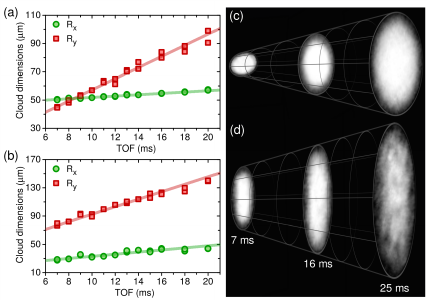}
\caption{(color online) BEC expansion during TOF. The Tomas-Fermi radii, $R_x$ and $R_y$, as a function of the TOF are show for standard (a) and turbulent BECs (b), respectively. (c), and (d) Three different TOF snapshots showing the expansion of standard and turbulent BECs, respectively.}
\label{fig:Fig1}
\end{figure}

	The typical results determined from the free fall expansion of the turbulent and the non-turbulent BECs are presented in Fig.~\ref{fig:Fig1}, where the clouds' typical sizes are plotted as a function of the elapsed time of flight. For a standard (non-turbulent) BEC, the expected aspect ratio inversion of the Thomas-Fermi radii during the free fall expansion takes place at about $8.5\unit{ms}$ TOF, as seen in Fig.~\ref{fig:Fig1}(a). On the other hand, the turbulent BECs are known to expand in a self-similar manner, without ever inverting its aspect ratio, shown in Fig.~\ref{fig:Fig1}(b). The turbulent cloud presents a self-similar anomalous expansion with a shape and an aspect ratio nearly constant during the TOF~\cite{Caracanhas2013,Caracanhas2012}. Essentially, the preferential vortex line alignment along the radial direction, transverse to the symmetry axis direction, balances the radii expansion rates keeping the aspect ratio locked during the TOF expansion~\cite{Caracanhas2013}.
The rate expansion determined in both cases are: $\dot{R_y} = 3.9(1)\unit{\mu m/ms}$, $\dot{R_x} = 0.47(1)\unit{\mu m/ms}$ for the standard BEC and $\dot{\tilde{R_y}} = 5.3(2)\unit{\mu m/ms}$, $\dot{\tilde{R_x}} = 1.5(1)\unit{\mu m/ms}$ for the turbulent BECs, clearly showing that the disordered matter-wave expands faster than the coherent one. Figures~\ref{fig:Fig1}(c) and \ref{fig:Fig1}(d) show the expansion of a standard (non-turbulent) BEC and a turbulent BEC, respectively, for different TOF intervals. Fig.~\ref{fig:Fig1}(d) presents the number density distribution and its characteristic modulations regularly observed in expanding turbulent BECs. Those can only be seen in detail on snapshots acquired after sufficiently long TOF intervals ($\geq25\unit{ms}$), when the optical depth lowers and the probe light passes through the atomic cloud revealing its typical density distribution pattern. This fact forces a lower limit on the TOF to allow for the experimental study of turbulent density distribution patterns. We shall mention that, currently, it is not technically possible to perform studies at TOFs longer than $55\unit{ms}$ in our system.

\begin{figure}
\centering
\includegraphics[width=\linewidth]{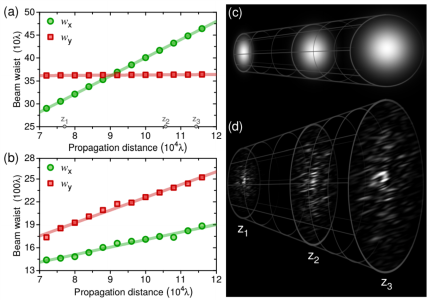}	
\caption{(color online) Optical beam propagation. The optical beam waists along the $x$ and $y$ directions, $w_x$ and $w_y$, as a function of the axial distance for a coherent Gaussian beam (a) and the speckle beam (b), respectively. The marked propagation distances $z_1$, $z_2$ and $z_3$, the Gaussian beam aspect ratio matches that of a standard BEC for  three TOFs in the Fig.~\ref{fig:Fig1}(c). (c), (d) Sequence of three different propagation distances showing the expansion of a Gaussian beam and a speckle beam, respectively.}
\label{fig:Fig2}
\end{figure}

To follow up the observations for the propagating matter wave, the propagation of light beams with and without speckle is presented in figure~\ref{fig:Fig2}. A coherent elliptical Gaussian laser beam propagates in space presenting aspect ratio inversion along the optical axis, similarly to the aspect ratio inversion observed in coherent matter waves, such as a free falling BEC. In that case, the traveling optical wave is characterized by its divergence angle, given by~\cite{Kogelnik1966}: $\tan\theta_i = {\lambda}/{\pi w_i}$, where $\lambda$ is the light wavelength and $w_i$ the beam waist in the $i$-direction. If we now consider the same elliptical beam, but with a speckle pattern on it, the divergence is quite different. The divergence angle for each direction is given by~\cite{Giglio2000,GoodmanBook2000} $\tan\theta_i = {\lambda}/{\ell_i}$, where $\ell_i$ is the speckle correlation length in the $i$-direction. Actually, $\ell_i$ is normally expressed as $d$ which is the typical particle size found in the rough surface that generated the speckle pattern. For an isotropic speckle field the correlation length is equal in both directions, thus the divergence is the same in any direction. 
	
	When dealing with light wave propagation one has to take into account different regimes, namely the near and the far field. The near field propagation regime takes place at short distances, compared to the Rayleigh length, and each and every domain expands as $\lambda/\ell_i$, which is precisely the case studied here. The overall expansion is given by the sum of individual domains, and therefore the general spatial evolution takes place described based on different expansion rates, for each direction, since there are different number in each direction. The propagation does not present the aspect ratio inversion, which tends to unity at large distances (far-field). In Fig.~\ref{fig:Fig2}, the axial evolution of the light beam waists of a coherent Gaussian beam and a speckle beam are shown as a function of the propagation distance from the focus. We point out that both coherent and incoherent optical beams start with identical elliptical cross-sections, but evolve at distinct rates due to the presence or lack of disorder (speckle). 
    
    For the coherent beam, Fig.~\ref{fig:Fig2}(a), the inversion takes place after a given propagation distance away from the focus. In contrast, for the speckle field, Fig.~\ref{fig:Fig2}(b), the beam propagates in a nearly self-similar shape, and the cross-section along each dimension increases faster when compared with the coherent beam. That happens when $\ell_i < w_i$, i.e. for a beam full of speckle grains, and presents a larger divergence angle. From figures~\ref{fig:Fig2}(a) and \ref{fig:Fig2}(b) the determined beam divergence in both cases are $\dot{w}_x = 3.964(2)\times10^{-2}$, $\dot{w}_y = 4.8(1)\times10^{-4}$ for the coherent Gaussian beam and $\dot{\tilde{w}}_x = 9.8(5)\times10^{-2}$ and $\dot{\tilde{w}}_y = 1.67(6)\times10^{-1}$ for the speckle beam, showing that the disordered light wave expands faster than the coherent one, just as the turbulent quantum gas, presented in Fig.~\ref{fig:Fig1}. Figures~\ref{fig:Fig2}(c) and (d) show the propagation of the coherent Gaussian and speckle beams, respectively. Those properties presented in Fig.~\ref{fig:Fig1} and Fig.~\ref{fig:Fig2} show the equivalence between the propagating speckle light field and the expansion of the corresponding matter wave originated from the atomic BEC strongly driven out-of-equilibrium. Both situations are equivalent for propagation regimes within the Rayleigh range. For much longer time propagation, both waves shall evolve to  isotropic expansion, asymptotically. We have verified this behavior by numerical simulations for the case of a light beam with speckle.

	Next, we evaluate the density-density correlation function for the expanded matter wave and compare it to that of an optical speckle, whose coherence (or the lack of it) can be expressed by the intensity-intensity correlation function. That was originally established by Hanbury-Brown-Twiss in their quest for the determination of the typical sizes of astronomical objects \cite{Hanbury1956}.
	
	The density function, $\rho(r)$, presents fluctuations due to turbulence in the sample. It is convenient to extract from $\rho(r)$ an average component and a fluctuation, chaotic component, $\rho(r) =  \langle \rho(r)\rangle + \rho_{fl}(r)$. By definition, the average of $\rho_{fl}(r)$ is zero, $\langle\rho_{fl}(r)\rangle = 0$. This property allows one to evaluate the density-density correlation function to study the the turbulent BEC, given by:
\begin{equation*}
C^{(2)}(r - r^{\prime})=\frac{\left\langle\rho(r)~\rho(r^{\prime})\right\rangle}{\left\langle\rho(r)\right\rangle \left\langle\rho(r^{\prime})\right\rangle} - 1 \\ 
					=\frac{\left\langle\rho_{fl}(r)~\rho_{fl}(r^{\prime})\right\rangle}{\left\langle\rho(r)\right\rangle \left\langle\rho(r^{\prime})\right\rangle},  \label{eq:Corr}
\end{equation*}
and thus, according to the above definition, \eqref{eq:Corr}, the correlation function for a standard BEC is null.

	Concerning the fluctuations observed on the atom number density, $\rho_{fl}(r)$, the probability distribution, $P(\rho_{fl})$, is expected to be a normalized Gaussian, as a result of the Central Limit Theorem. Thus $P(\rho_{fl})=N\exp{[-(\rho_{fl})^2/2\sigma^2]}$, where $\sigma^2$ is the variance, $\sigma^2 = \langle( \rho_{fl})^2\rangle$, and $N$ is the normalization constant, $N = \sqrt{2/\pi\sigma^2}$. The variance is basically the correlation function at $\Delta r=r-r^{\prime}=0$. It is convenient to present our result for the correlation function normalized to the variance, such that at $\Delta r = 0$ the correlation function is unity. In this way, we define the normalized correlation function as $\widetilde{C}^{(2)}(\Delta r) = C^{(2)}(\Delta r)/C^{(2)}(0)$. Of course in order to calculate averages such as $\langle \rho_{fl}(r) \rho_{fl}(r ^{\prime})\rangle$, one needs the functional $r$-dependence of $\rho_{fl}(r)$. Not having this at hand, we merely construct the average through sampling different values of $\Delta r$, and then performing the average as $\sum_{i = 1}^{M} \rho_{fl}(r_i) \rho_{fl}(r ^{\prime}_i)/M$. The large $\Delta r$ behavior of the correlation function is expected to be a Gaussian or an exponential. In our analysis to follow the exponential decay with increasing $\Delta r$ was found to better represent the behavior of the data, and the correlation length can be extracted by using $\widetilde{C}^{(2)}(\ell) = 1/e$. In the following we calculate the correlation function using the measured atomic density, $\rho(r)$, obtained after expansion in time of flight.

\begin{figure}
\centering
\includegraphics[width=\linewidth]{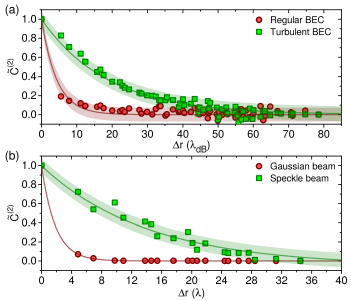}	
\caption{(color online) Normalized correlation function, $\widetilde{C}^{(2)}(\Delta r)$. (a) For a regular BEC (red) and a turbulent cloud (green) as a function of the thermal De Broglie wavelentgh, $\lambda_{dB}$. (b) For the numerical simulations of the a Gaussian (red) and a speckle beam of light (green) as a function of $\Delta{r}$, in units of $\lambda$, at a propagation distance where the Gaussian beam aspect ratio matches with the regular BEC of (a). The $\widetilde{C}^{(2)}(\Delta r)$ for a regular BEC (Gaussian beam) is zero except for very small $\Delta r$, whereas for a turbulent cloud (speckle beam) it has an exponential decay with a correlation length of $\ell_{Turb} = 8.5(2)~\mu$m ($\ell_{Speckle} = 13.8(8)\lambda$). The typical dimensions are $R_x = 63(3)~\mu$m and $R_y = 198(6)~\mu$m for the BEC; and $w_x = 1876\lambda$ and $w_y = 2528\lambda$ for the beam waists, showing that the correlation length is much smaller than the BEC/beam sizes. The colored regions represent the confidence bands of the exponential decay fittings.}
\label{fig:Fig3}
\end{figure}

        In Fig.~\ref{fig:Fig3}(a), we analyze the normalized correlation function $\widetilde{C}^{(2)}(\Delta r)$ of both: a regular and a strongly perturbed BEC, after $30\unit{ms}$ of free fall. The correlation for a regular BEC is basically constant and close to zero. At short distances, $\widetilde{C}^{(2)}(\Delta{r})$ for the regular BEC is greater than zero due to the small thermal fluctuations originated from the thermal atoms component of the sample. The turbulent sample shows a typical $\widetilde{C}^{(2)}(\Delta{r})$ of a normalized decaying form at larger $\Delta r$, which is a result of a much shorter correlation length ($\ell_{Turb}=8.5(2)~\unit{\mu m}$) than the sizes of the cloud ($R_x = 63(3)~\unit{\mu m}$ and $R_y = 198(6)~\unit{\mu m}$) and indicates the presence of fluctuations at smaller scales. The in-trap correlation length can be extracted from the TOF correlation length by $\ell_{Turb}^{0} = \ell_{Turb}/\sqrt{1+(\omega_{r}t)^2} \approx 1.3\xi$ \cite{Castin1996}, where $t$ is the TOF and $\xi = 0.18~\unit{\mu m}$ is the BEC healing length, which is also the typical vortex radius. The correlation length clearly indicates the dynamical randomness of the non-equilibrium BEC is entirely connected to the vortices originally created in the disturbed BEC, with further developing.

\begin{figure}
\centering
\includegraphics[width=\linewidth]{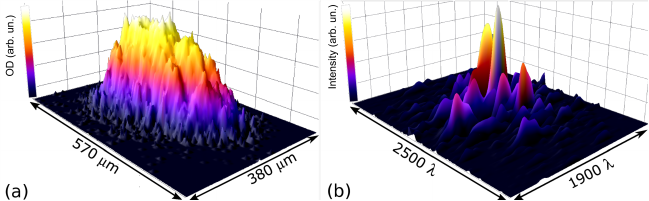}	
\caption{\label{Fig3}(color online) Direct comparison in between a turbulent BEC and an elliptical speckle beam. (a) Typical atom number density map of a turbulent condensate shown in units of optical depth (OD). (b) Light intensity map from a cross section of a speckle beam. The integration in one of the 3D dimension of the BEC results in slightly smaller amplitude for the density fluctuations.}
\label{fig:Fig4}
\end{figure}

	Such a behavior is very similar to a speckle light field, which motivated us to look for an underlying analogy existing in between the light field and the expanding quantum degenerate gas. Indeed, the speckle field arises from a sum of complex statistically independent amplitudes bearing a random magnitude and phase~\cite{Goodman2010Book}. For this purpose, we simulated a propagating coherent Gaussian beam and speckle light beam and compare with the expanding BEC and non-equilibrium BEC. 
 In Fig.~\ref{fig:Fig4}(b), we present the normalized correlation function $\widetilde{C}^{(2)}(\Delta r)$ of a Gaussian and a speckle beam at a propagation distance where the Gaussian beam aspect ratio matches that of a regular BEC of Fig.~\ref{fig:Fig4}(a). Essentially, the same qualitative decay behavior of the correlation function is observed on the light field and on the expanding quantum gas. In Fig.~\ref{fig:Fig4}, we present the atomic density absorption profile of an expanding turbulent cloud together with the cross section intensity profile of a propagating speckle field, showing the similarities between the distribution of the spatial fluctuations amplitudes. An important point concerning our measurements is that they are obtained from absorption images. This corresponds to a projection of the condensate's atomic (number) density distribution onto an image plane. Despite the fact that this procedure smooths the fluctuations amplitude (as shown in Fig.~\ref{fig:Fig4}), the correlation length measured after projection remains the same, as in three dimensions. This is supported by 2D simulations of an optical speckle beam and its unidimensional projection. It was observed that the effect on the correlation length, extracted  from $\widetilde{C}^{(2)}(\Delta{r})$, remains unaffected. In fact, the observation of quite a similar behavior found in between the spatial evolution of such different physical systems, starting from completely different initial conditions, is truly remarkable.
        
	The observation and study of disorder existing in matter waves freely moving in space, which can be accomplished by driving Bose-Einstein condensates to turbulence, may open up a broad new research field with opportunities for understanding of the inner nature of the disorder phenomena. In particular, we mention the spatial propagation of matter-waves speckle. We note that the interaction of matter-wave with random potentials created by speckle light fields has already produced very interesting results on Anderson localization \cite{Sanchez-Palencia2007,Billy2008,Roati2008}. Finally, we point out that there might also be different means of nucleating disorder into quantum gases (e.g., via modulation of microscopic interactions). We do believe that similar results may be achieved once disorder in amplitude and phase is established, though we have only explored it via the turbulent regime in this work.

	In conclusion, we have investigated the expansion of out-of-equilibrium (turbulent) BECs and found strong evidences that it evolves in space much like a matter-wave speckle field. This finding may well be the possibility to investigate the statistical properties of matter waves. The speckle-like behavior observed in free expanding quantum fluids, driven out-of-equilibrium, may emerge as an ideal candidate system to explore disordered patterns and the details of speckle fields traveling in three dimensions.

\section*{Significance}
In recent years, significant progress concerning matter-wave creation and manipulation has been achieved. Fresnel diffraction, atom lasers, and nonlinear matter wave phenomena, are just a few examples of recent interesting research topics. There have also been quite a lot of studies related to the propagation of matter waves interacting with disordered potentials, but not as much concerning the disorder existing into the matter wave itself. In the present study, we analyze the disorder imprinted into a Bose-Einstein condensate (BEC), and its characteristic spatial evolution afterwards. We performed careful analyses and traced a few analogies that took us to the conclusion that the perturbed BECs evolve very much in a speckle-like manner, in close analogy with a traveling light (optical) speckle.

\begin{acknowledgments}
We thank FAPESP (grant 2013/07276-1), CAPES and CNPq (grant PVE 400228/2014-9) for financial support. We appreciate the discussions with P. Courteille (IFSC), and the technical assistance provided by E. Henn (IFSC) and G. Roati (LENS/Italy). MSH also acknowledges a Senior Visiting Professorship granted by CAPES (program CAPES/ITA-PVS).
\end{acknowledgments}

\bibliographystyle{apsrev4-1}
\bibliography{Speckle-References}

\begin{thebibliography}{21}%
\makeatletter
\providecommand \@ifxundefined [1]{%
 \@ifx{#1\undefined}
}%
\providecommand \@ifnum [1]{%
 \ifnum #1\expandafter \@firstoftwo
 \else \expandafter \@secondoftwo
 \fi
}%
\providecommand \@ifx [1]{%
 \ifx #1\expandafter \@firstoftwo
 \else \expandafter \@secondoftwo
 \fi
}%
\providecommand \natexlab [1]{#1}%
\providecommand \enquote  [1]{``#1''}%
\providecommand \bibnamefont  [1]{#1}%
\providecommand \bibfnamefont [1]{#1}%
\providecommand \citenamefont [1]{#1}%
\providecommand \href@noop [0]{\@secondoftwo}%
\providecommand \href [0]{\begingroup \@sanitize@url \@href}%
\providecommand \@href[1]{\@@startlink{#1}\@@href}%
\providecommand \@@href[1]{\endgroup#1\@@endlink}%
\providecommand \@sanitize@url [0]{\catcode `\\12\catcode `\$12\catcode
  `\&12\catcode `\#12\catcode `\^12\catcode `\_12\catcode `\%12\relax}%
\providecommand \@@startlink[1]{}%
\providecommand \@@endlink[0]{}%
\providecommand \url  [0]{\begingroup\@sanitize@url \@url }%
\providecommand \@url [1]{\endgroup\@href {#1}{\urlprefix }}%
\providecommand \urlprefix  [0]{URL }%
\providecommand \Eprint [0]{\href }%
\providecommand \doibase [0]{http://dx.doi.org/}%
\providecommand \selectlanguage [0]{\@gobble}%
\providecommand \bibinfo  [0]{\@secondoftwo}%
\providecommand \bibfield  [0]{\@secondoftwo}%
\providecommand \translation [1]{[#1]}%
\providecommand \BibitemOpen [0]{}%
\providecommand \bibitemStop [0]{}%
\providecommand \bibitemNoStop [0]{.\EOS\space}%
\providecommand \EOS [0]{\spacefactor3000\relax}%
\providecommand \BibitemShut  [1]{\csname bibitem#1\endcsname}%
\let\auto@bib@innerbib\@empty
\bibitem [{\citenamefont {Castin}\ and\ \citenamefont
  {Dum}(1996)}]{Castin1996}%
  \BibitemOpen
  \bibfield  {author} {\bibinfo {author} {\bibfnamefont {Y.}~\bibnamefont
  {Castin}}\ and\ \bibinfo {author} {\bibfnamefont {R.}~\bibnamefont {Dum}},\
  }\href {\doibase 10.1103/PhysRevLett.77.5315} {\bibfield  {journal} {\bibinfo
   {journal} {Phys. Rev. Lett.}\ }\textbf {\bibinfo {volume} {77}},\ \bibinfo
  {pages} {5315} (\bibinfo {year} {1996})}\BibitemShut {NoStop}%
\bibitem [{\citenamefont {Impens}\ and\ \citenamefont
  {Bord\'e}(2009)}]{Impens2009}%
  \BibitemOpen
  \bibfield  {author} {\bibinfo {author} {\bibfnamefont {F.}~\bibnamefont
  {Impens}}\ and\ \bibinfo {author} {\bibfnamefont {C.~J.}\ \bibnamefont
  {Bord\'e}},\ }\href {\doibase 10.1103/PhysRevA.79.043613} {\bibfield
  {journal} {\bibinfo  {journal} {Phys. Rev. A}\ }\textbf {\bibinfo {volume}
  {79}},\ \bibinfo {pages} {043613} (\bibinfo {year} {2009})}\BibitemShut
  {NoStop}%
\bibitem [{\citenamefont {Riou}\ \emph {et~al.}(2008)\citenamefont {Riou},
  \citenamefont {Le~Coq}, \citenamefont {Impens}, \citenamefont {Guerin},
  \citenamefont {Bord\'e}, \citenamefont {Aspect},\ and\ \citenamefont
  {Bouyer}}]{Riou2008}%
  \BibitemOpen
  \bibfield  {author} {\bibinfo {author} {\bibfnamefont {J.~F.}\ \bibnamefont
  {Riou}}, \bibinfo {author} {\bibfnamefont {Y.}~\bibnamefont {Le~Coq}},
  \bibinfo {author} {\bibfnamefont {F.}~\bibnamefont {Impens}}, \bibinfo
  {author} {\bibfnamefont {W.}~\bibnamefont {Guerin}}, \bibinfo {author}
  {\bibfnamefont {C.~J.}\ \bibnamefont {Bord\'e}}, \bibinfo {author}
  {\bibfnamefont {A.}~\bibnamefont {Aspect}}, \ and\ \bibinfo {author}
  {\bibfnamefont {P.}~\bibnamefont {Bouyer}},\ }\href {\doibase
  10.1103/PhysRevA.77.033630} {\bibfield  {journal} {\bibinfo  {journal} {Phys.
  Rev. A}\ }\textbf {\bibinfo {volume} {77}},\ \bibinfo {pages} {033630}
  (\bibinfo {year} {2008})}\BibitemShut {NoStop}%
\bibitem [{\citenamefont {Le~Coq}\ \emph {et~al.}(2001)\citenamefont {Le~Coq},
  \citenamefont {Thywissen}, \citenamefont {Rangwala}, \citenamefont {Gerbier},
  \citenamefont {Richard}, \citenamefont {Delannoy}, \citenamefont {Bouyer},\
  and\ \citenamefont {Aspect}}]{Le2001}%
  \BibitemOpen
  \bibfield  {author} {\bibinfo {author} {\bibfnamefont {Y.}~\bibnamefont
  {Le~Coq}}, \bibinfo {author} {\bibfnamefont {J.~H.}\ \bibnamefont
  {Thywissen}}, \bibinfo {author} {\bibfnamefont {S.~A.}\ \bibnamefont
  {Rangwala}}, \bibinfo {author} {\bibfnamefont {F.}~\bibnamefont {Gerbier}},
  \bibinfo {author} {\bibfnamefont {S.}~\bibnamefont {Richard}}, \bibinfo
  {author} {\bibfnamefont {G.}~\bibnamefont {Delannoy}}, \bibinfo {author}
  {\bibfnamefont {P.}~\bibnamefont {Bouyer}}, \ and\ \bibinfo {author}
  {\bibfnamefont {A.}~\bibnamefont {Aspect}},\ }\href {\doibase
  10.1103/PhysRevLett.87.170403} {\bibfield  {journal} {\bibinfo  {journal}
  {Phys. Rev. Lett.}\ }\textbf {\bibinfo {volume} {87}},\ \bibinfo {pages}
  {170403} (\bibinfo {year} {2001})}\BibitemShut {NoStop}%
\bibitem [{\citenamefont {Busch}\ \emph {et~al.}(2002)\citenamefont {Busch},
  \citenamefont {K\"ohl}, \citenamefont {Esslinger},\ and\ \citenamefont
  {M\o{}lmer}}]{Busch2002}%
  \BibitemOpen
  \bibfield  {author} {\bibinfo {author} {\bibfnamefont {T.}~\bibnamefont
  {Busch}}, \bibinfo {author} {\bibfnamefont {M.}~\bibnamefont {K\"ohl}},
  \bibinfo {author} {\bibfnamefont {T.}~\bibnamefont {Esslinger}}, \ and\
  \bibinfo {author} {\bibfnamefont {K.}~\bibnamefont {M\o{}lmer}},\ }\href
  {\doibase 10.1103/PhysRevA.65.043615} {\bibfield  {journal} {\bibinfo
  {journal} {Phys. Rev. A}\ }\textbf {\bibinfo {volume} {65}},\ \bibinfo
  {pages} {043615} (\bibinfo {year} {2002})}\BibitemShut {NoStop}%
\bibitem [{\citenamefont {Riou}\ \emph {et~al.}(2006)\citenamefont {Riou},
  \citenamefont {Guerin}, \citenamefont {Coq}, \citenamefont {Fauquembergue},
  \citenamefont {Josse}, \citenamefont {Bouyer},\ and\ \citenamefont
  {Aspect}}]{Riou2006}%
  \BibitemOpen
  \bibfield  {author} {\bibinfo {author} {\bibfnamefont {J.~F.}\ \bibnamefont
  {Riou}}, \bibinfo {author} {\bibfnamefont {W.}~\bibnamefont {Guerin}},
  \bibinfo {author} {\bibfnamefont {Y.~L.}\ \bibnamefont {Coq}}, \bibinfo
  {author} {\bibfnamefont {M.}~\bibnamefont {Fauquembergue}}, \bibinfo {author}
  {\bibfnamefont {V.}~\bibnamefont {Josse}}, \bibinfo {author} {\bibfnamefont
  {P.}~\bibnamefont {Bouyer}}, \ and\ \bibinfo {author} {\bibfnamefont
  {A.}~\bibnamefont {Aspect}},\ }\href {\doibase 10.1103/PhysRevLett.96.070404}
  {\bibfield  {journal} {\bibinfo  {journal} {Phys. Rev. Lett.}\ }\textbf
  {\bibinfo {volume} {96}},\ \bibinfo {pages} {070404} (\bibinfo {year}
  {2006})}\BibitemShut {NoStop}%
\bibitem [{\citenamefont {Couvert}\ \emph {et~al.}(2008)\citenamefont
  {Couvert}, \citenamefont {Jeppesen}, \citenamefont {Kawalec}, \citenamefont
  {Reinaudi}, \citenamefont {Mathevet},\ and\ \citenamefont
  {Gu\'ery-Odelin}}]{Couvert2008}%
  \BibitemOpen
  \bibfield  {author} {\bibinfo {author} {\bibfnamefont {A.}~\bibnamefont
  {Couvert}}, \bibinfo {author} {\bibfnamefont {M.}~\bibnamefont {Jeppesen}},
  \bibinfo {author} {\bibfnamefont {T.}~\bibnamefont {Kawalec}}, \bibinfo
  {author} {\bibfnamefont {G.}~\bibnamefont {Reinaudi}}, \bibinfo {author}
  {\bibfnamefont {R.}~\bibnamefont {Mathevet}}, \ and\ \bibinfo {author}
  {\bibfnamefont {D.}~\bibnamefont {Gu\'ery-Odelin}},\ }\href {\doibase
  10.1209/0295-5075/83/50001} {\bibfield  {journal} {\bibinfo  {journal}
  {Europhys. Lett.}\ }\textbf {\bibinfo {volume} {83}},\ \bibinfo {pages}
  {50001} (\bibinfo {year} {2008})}\BibitemShut {NoStop}%
\bibitem [{\citenamefont {Goodman}(2010)}]{Goodman2010Book}%
  \BibitemOpen
  \bibfield  {author} {\bibinfo {author} {\bibfnamefont {J.~W.}\ \bibnamefont
  {Goodman}},\ }\href {\doibase 10.1007/s10955-007-9440-8} {\emph {\bibinfo
  {title} {Speckle Phenomena in Optics: Theory and Applications}}}\ (\bibinfo
  {publisher} {Roberts \& Company},\ \bibinfo {year} {2010})\BibitemShut
  {NoStop}%
\bibitem [{\citenamefont {Patton}\ \emph {et~al.}(2006)\citenamefont {Patton},
  \citenamefont {Deponte}, \citenamefont {Elliott},\ and\ \citenamefont
  {Kevan}}]{Patton2006}%
  \BibitemOpen
  \bibfield  {author} {\bibinfo {author} {\bibfnamefont {F.~S.}\ \bibnamefont
  {Patton}}, \bibinfo {author} {\bibfnamefont {D.~P.}\ \bibnamefont {Deponte}},
  \bibinfo {author} {\bibfnamefont {G.~S.}\ \bibnamefont {Elliott}}, \ and\
  \bibinfo {author} {\bibfnamefont {S.~D.}\ \bibnamefont {Kevan}},\ }\href
  {\doibase 10.1103/PhysRevLett.97.013202} {\bibfield  {journal} {\bibinfo
  {journal} {Phys. Rev. Lett.}\ }\textbf {\bibinfo {volume} {97}},\ \bibinfo
  {pages} {013202} (\bibinfo {year} {2006})}\BibitemShut {NoStop}%
\bibitem [{\citenamefont {Dall}\ \emph {et~al.}(2011)\citenamefont {Dall},
  \citenamefont {Hodgman}, \citenamefont {Manning}, \citenamefont {Johnsson},
  \citenamefont {Baldwin},\ and\ \citenamefont {Truscott}}]{Dall2011}%
  \BibitemOpen
  \bibfield  {author} {\bibinfo {author} {\bibfnamefont {R.~G.}\ \bibnamefont
  {Dall}}, \bibinfo {author} {\bibfnamefont {S.~S.}\ \bibnamefont {Hodgman}},
  \bibinfo {author} {\bibfnamefont {A.~G.}\ \bibnamefont {Manning}}, \bibinfo
  {author} {\bibfnamefont {M.~T.}\ \bibnamefont {Johnsson}}, \bibinfo {author}
  {\bibfnamefont {K.~G.~H.}\ \bibnamefont {Baldwin}}, \ and\ \bibinfo {author}
  {\bibfnamefont {A.~G.}\ \bibnamefont {Truscott}},\ }\href {\doibase
  10.1038/ncomms1292} {\bibfield  {journal} {\bibinfo  {journal} {Nat. Comm.}\
  }\textbf {\bibinfo {volume} {2}},\ \bibinfo {pages} {291} (\bibinfo {year}
  {2011})}\BibitemShut {NoStop}%
\bibitem [{\citenamefont {Thompson}\ \emph {et~al.}(2014)\citenamefont
  {Thompson}, \citenamefont {Bagnato}, \citenamefont {Telles}, \citenamefont
  {Caracanhas}, \citenamefont {dos Santos},\ and\ \citenamefont
  {Bagnato}}]{Thompson2014}%
  \BibitemOpen
  \bibfield  {author} {\bibinfo {author} {\bibfnamefont {K.~J.}\ \bibnamefont
  {Thompson}}, \bibinfo {author} {\bibfnamefont {G.~G.}\ \bibnamefont
  {Bagnato}}, \bibinfo {author} {\bibfnamefont {G.~D.}\ \bibnamefont {Telles}},
  \bibinfo {author} {\bibfnamefont {M.~A.}\ \bibnamefont {Caracanhas}},
  \bibinfo {author} {\bibfnamefont {F.~E.~A.}\ \bibnamefont {dos Santos}}, \
  and\ \bibinfo {author} {\bibfnamefont {V.~S.}\ \bibnamefont {Bagnato}},\
  }\href {\doibase 10.1088/1612-2011/11/1/015501} {\bibfield  {journal}
  {\bibinfo  {journal} {Laser Phys. Lett.}\ }\textbf {\bibinfo {volume} {11}},\
  \bibinfo {pages} {015501} (\bibinfo {year} {2014})}\BibitemShut {NoStop}%
\bibitem [{\citenamefont {Tsatsos}\ \emph {et~al.}(2016)\citenamefont
  {Tsatsos}, \citenamefont {Tavares}, \citenamefont {Cidrim}, \citenamefont
  {Fritsch}, \citenamefont {Caracanhas}, \citenamefont {dos Santos},
  \citenamefont {Barenghi},\ and\ \citenamefont {Bagnato}}]{Tsatsos2016}%
  \BibitemOpen
  \bibfield  {author} {\bibinfo {author} {\bibfnamefont {M.~C.}\ \bibnamefont
  {Tsatsos}}, \bibinfo {author} {\bibfnamefont {P.~E.~S.}\ \bibnamefont
  {Tavares}}, \bibinfo {author} {\bibfnamefont {A.}~\bibnamefont {Cidrim}},
  \bibinfo {author} {\bibfnamefont {A.~R.}\ \bibnamefont {Fritsch}}, \bibinfo
  {author} {\bibfnamefont {M.~A.}\ \bibnamefont {Caracanhas}}, \bibinfo
  {author} {\bibfnamefont {F.~E.~A.}\ \bibnamefont {dos Santos}}, \bibinfo
  {author} {\bibfnamefont {C.~F.}\ \bibnamefont {Barenghi}}, \ and\ \bibinfo
  {author} {\bibfnamefont {V.~S.}\ \bibnamefont {Bagnato}},\ }\href {\doibase
  10.1016/j.physrep.2016.02.003} {\bibfield  {journal} {\bibinfo  {journal}
  {Physics Reports}\ }\textbf {\bibinfo {volume} {622}},\ \bibinfo {pages} {1}
  (\bibinfo {year} {2016})}\BibitemShut {NoStop}%
\bibitem [{\citenamefont {Caracanhas}\ \emph {et~al.}(2013)\citenamefont
  {Caracanhas}, \citenamefont {Fetter}, \citenamefont {Baym}, \citenamefont
  {Muniz},\ and\ \citenamefont {Bagnato}}]{Caracanhas2013}%
  \BibitemOpen
  \bibfield  {author} {\bibinfo {author} {\bibfnamefont {M.}~\bibnamefont
  {Caracanhas}}, \bibinfo {author} {\bibfnamefont {A.~L.}\ \bibnamefont
  {Fetter}}, \bibinfo {author} {\bibfnamefont {G.}~\bibnamefont {Baym}},
  \bibinfo {author} {\bibfnamefont {S.~R.}\ \bibnamefont {Muniz}}, \ and\
  \bibinfo {author} {\bibfnamefont {V.~S.}\ \bibnamefont {Bagnato}},\ }\href
  {\doibase 10.1007/s10909-012-0776-3} {\bibfield  {journal} {\bibinfo
  {journal} {J. Low Temp. Phys.}\ }\textbf {\bibinfo {volume} {170}},\ \bibinfo
  {pages} {133} (\bibinfo {year} {2013})}\BibitemShut {NoStop}%
\bibitem [{\citenamefont {Caracanhas}\ \emph {et~al.}(2012)\citenamefont
  {Caracanhas}, \citenamefont {Fetter}, \citenamefont {Muniz}, \citenamefont
  {Magalh{\~a}es}, \citenamefont {Roati}, \citenamefont {Bagnato},\ and\
  \citenamefont {Bagnato}}]{Caracanhas2012}%
  \BibitemOpen
  \bibfield  {author} {\bibinfo {author} {\bibfnamefont {M.}~\bibnamefont
  {Caracanhas}}, \bibinfo {author} {\bibfnamefont {A.}~\bibnamefont {Fetter}},
  \bibinfo {author} {\bibfnamefont {S.}~\bibnamefont {Muniz}}, \bibinfo
  {author} {\bibfnamefont {K.}~\bibnamefont {Magalh{\~a}es}}, \bibinfo {author}
  {\bibfnamefont {G.}~\bibnamefont {Roati}}, \bibinfo {author} {\bibfnamefont
  {G.}~\bibnamefont {Bagnato}}, \ and\ \bibinfo {author} {\bibfnamefont
  {V.}~\bibnamefont {Bagnato}},\ }\href {\doibase 10.1007/s10909-011-0409-2}
  {\bibfield  {journal} {\bibinfo  {journal} {J. Low Temp. Phys.}\ }\textbf
  {\bibinfo {volume} {166}},\ \bibinfo {pages} {49} (\bibinfo {year}
  {2012})}\BibitemShut {NoStop}%
\bibitem [{\citenamefont {Kogelnik}\ and\ \citenamefont
  {Li}(1966)}]{Kogelnik1966}%
  \BibitemOpen
  \bibfield  {author} {\bibinfo {author} {\bibfnamefont {H.}~\bibnamefont
  {Kogelnik}}\ and\ \bibinfo {author} {\bibfnamefont {T.}~\bibnamefont {Li}},\
  }\href {\doibase 10.1109/PROC.1966.5119} {\bibfield  {journal} {\bibinfo
  {journal} {Proceedings of the IEEE}\ }\textbf {\bibinfo {volume} {54}},\
  \bibinfo {pages} {1312} (\bibinfo {year} {1966})}\BibitemShut {NoStop}%
\bibitem [{\citenamefont {Giglio}\ \emph {et~al.}(2000)\citenamefont {Giglio},
  \citenamefont {Carpineti},\ and\ \citenamefont {Vailati}}]{Giglio2000}%
  \BibitemOpen
  \bibfield  {author} {\bibinfo {author} {\bibfnamefont {M.}~\bibnamefont
  {Giglio}}, \bibinfo {author} {\bibfnamefont {M.}~\bibnamefont {Carpineti}}, \
  and\ \bibinfo {author} {\bibfnamefont {A.}~\bibnamefont {Vailati}},\ }\href
  {\doibase 10.1103/PhysRevLett.85.1416} {\bibfield  {journal} {\bibinfo
  {journal} {Phys. Rev. Lett.}\ }\textbf {\bibinfo {volume} {85}},\ \bibinfo
  {pages} {1416} (\bibinfo {year} {2000})}\BibitemShut {NoStop}%
\bibitem [{\citenamefont {Goodman}(2000)}]{GoodmanBook2000}%
  \BibitemOpen
  \bibfield  {author} {\bibinfo {author} {\bibfnamefont {J.~W.}\ \bibnamefont
  {Goodman}},\ }\href {\doibase 10.1016/S0143-8166(01)00027-6} {\emph {\bibinfo
  {title} {Statistical Optics}}}\ (\bibinfo  {publisher} {John Wiley \& Sons},\
  \bibinfo {year} {2000})\BibitemShut {NoStop}%
\bibitem [{\citenamefont {Hanbury-Brown}\ and\ \citenamefont
  {Twiss}(1956)}]{Hanbury1956}%
  \BibitemOpen
  \bibfield  {author} {\bibinfo {author} {\bibfnamefont {R.}~\bibnamefont
  {Hanbury-Brown}}\ and\ \bibinfo {author} {\bibfnamefont {R.~Q.}\ \bibnamefont
  {Twiss}},\ }\href {\doibase 10.1038/177027a0} {\bibfield  {journal} {\bibinfo
   {journal} {Nature}\ ,\ \bibinfo {pages} {27}} (\bibinfo {year}
  {1956})}\BibitemShut {NoStop}%
\bibitem [{\citenamefont {Sanchez-Palencia}\ \emph {et~al.}(2007)\citenamefont
  {Sanchez-Palencia}, \citenamefont {Cl\'ement}, \citenamefont {Lugan},
  \citenamefont {Bouyer}, \citenamefont {Shlyapnikov},\ and\ \citenamefont
  {Aspect}}]{Sanchez-Palencia2007}%
  \BibitemOpen
  \bibfield  {author} {\bibinfo {author} {\bibfnamefont {L.}~\bibnamefont
  {Sanchez-Palencia}}, \bibinfo {author} {\bibfnamefont {D.}~\bibnamefont
  {Cl\'ement}}, \bibinfo {author} {\bibfnamefont {P.}~\bibnamefont {Lugan}},
  \bibinfo {author} {\bibfnamefont {P.}~\bibnamefont {Bouyer}}, \bibinfo
  {author} {\bibfnamefont {G.~V.}\ \bibnamefont {Shlyapnikov}}, \ and\ \bibinfo
  {author} {\bibfnamefont {A.}~\bibnamefont {Aspect}},\ }\href {\doibase
  10.1103/PhysRevLett.98.210401} {\bibfield  {journal} {\bibinfo  {journal}
  {Phys. Rev. Lett.}\ }\textbf {\bibinfo {volume} {98}},\ \bibinfo {pages}
  {210401} (\bibinfo {year} {2007})}\BibitemShut {NoStop}%
\bibitem [{\citenamefont {Billy}\ \emph {et~al.}(2008)\citenamefont {Billy},
  \citenamefont {Josse}, \citenamefont {Zuo}, \citenamefont {Bernard},
  \citenamefont {Hambrecht}, \citenamefont {Lugan}, \citenamefont {Cl\'ement},
  \citenamefont {Sanchez-Palencia}, \citenamefont {Bouyer},\ and\ \citenamefont
  {Aspect}}]{Billy2008}%
  \BibitemOpen
  \bibfield  {author} {\bibinfo {author} {\bibfnamefont {J.}~\bibnamefont
  {Billy}}, \bibinfo {author} {\bibfnamefont {V.}~\bibnamefont {Josse}},
  \bibinfo {author} {\bibfnamefont {Z.}~\bibnamefont {Zuo}}, \bibinfo {author}
  {\bibfnamefont {A.}~\bibnamefont {Bernard}}, \bibinfo {author} {\bibfnamefont
  {B.}~\bibnamefont {Hambrecht}}, \bibinfo {author} {\bibfnamefont
  {P.}~\bibnamefont {Lugan}}, \bibinfo {author} {\bibfnamefont
  {D.}~\bibnamefont {Cl\'ement}}, \bibinfo {author} {\bibfnamefont
  {L.}~\bibnamefont {Sanchez-Palencia}}, \bibinfo {author} {\bibfnamefont
  {P.}~\bibnamefont {Bouyer}}, \ and\ \bibinfo {author} {\bibfnamefont
  {A.}~\bibnamefont {Aspect}},\ }\href {\doibase 10.1038/nature07000}
  {\bibfield  {journal} {\bibinfo  {journal} {Nature}\ }\textbf {\bibinfo
  {volume} {453}},\ \bibinfo {pages} {891} (\bibinfo {year}
  {2008})}\BibitemShut {NoStop}%
\bibitem [{\citenamefont {Roati}\ \emph {et~al.}(2008)\citenamefont {Roati},
  \citenamefont {D'Errico}, \citenamefont {Fallani}, \citenamefont {Fattori},
  \citenamefont {Fort}, \citenamefont {Zaccanti}, \citenamefont {Modugno},
  \citenamefont {Modugno},\ and\ \citenamefont {Inguscio}}]{Roati2008}%
  \BibitemOpen
  \bibfield  {author} {\bibinfo {author} {\bibfnamefont {G.}~\bibnamefont
  {Roati}}, \bibinfo {author} {\bibfnamefont {C.}~\bibnamefont {D'Errico}},
  \bibinfo {author} {\bibfnamefont {L.}~\bibnamefont {Fallani}}, \bibinfo
  {author} {\bibfnamefont {M.}~\bibnamefont {Fattori}}, \bibinfo {author}
  {\bibfnamefont {C.}~\bibnamefont {Fort}}, \bibinfo {author} {\bibfnamefont
  {M.}~\bibnamefont {Zaccanti}}, \bibinfo {author} {\bibfnamefont
  {G.}~\bibnamefont {Modugno}}, \bibinfo {author} {\bibfnamefont
  {M.}~\bibnamefont {Modugno}}, \ and\ \bibinfo {author} {\bibfnamefont
  {M.}~\bibnamefont {Inguscio}},\ }\href {\doibase 10.1038/nature07071}
  {\bibfield  {journal} {\bibinfo  {journal} {Nature}\ }\textbf {\bibinfo
  {volume} {453}},\ \bibinfo {pages} {895} (\bibinfo {year}
  {2008})}\BibitemShut {NoStop}%
\end{thebibliography}%

\end{document}